\RequirePackage{fix-cm}
\documentclass[twocolumn,epjc3]{svjour3}  
\smartqed  % flush right qed marks, e.g. at end of proof
\RequirePackage{graphicx}
\RequirePackage{bm}
\RequirePackage{braket}
\RequirePackage{amsmath}
\journalname{Eur. Phys. J. A}
\begin{document}

\title{$\bm\alpha$ clustering and neutron-skin thickness of carbon isotopes}


\author{Q. Zhao\thanksref{addr1} 
        \and
        Y. Suzuki\thanksref{addr2}
        \and
        J. He\thanksref{addr2}
        \and
        B. Zhou\thanksref{addr2,addr3}
        \and
        M. Kimura\thanksref{e1,addr2,addr1,addr4}}

\thankstext{e1}{e-mail: masaaki@nucl.sci.hokdai.ac.jp}

\institute{Nuclear Reaction Data Centre (JCPRG), Hokkaido University, Sapporo 060-0810, Japan\label{addr1}
           \and
           Department of Physics, Hokkaido University, Sapporo 060-0810, Japan\label{addr2}
           \and
           Institute of Modern Physics, Fudan University, Shanghai 200433, China\label{addr3}
           \and
	   Research Center for Nuclear Physics (RCNP), Osaka University, Ibaraki 567-0047, Japan\label{addr4}
}

\date{Received: date / Accepted: date}

\maketitle

\begin{abstract}
 The interplay between the formation of neutron skin and $\alpha$ cluster at the dilute surface of
 neutron-rich nuclei is one of the interesting subjects in the study of neutron-rich nuclei and
 nuclear clustering. A theoretical model has predicted that the growth of neutron skin will prevent
 the  $\alpha$ clustering at nuclear surface.  Quite recently, this theoretical perspective; the
 suppression of $\alpha$ clustering by the  neutron-skin formation was firstly confirmed
 experimentally in Sn isotopes as the reduction of the  $(p,p\alpha)$  reaction cross section.
 Motivated by the novel discovery, in this work, we have   investigated the relationship between the
 neutron-skin thickness and $\alpha$ clustering in C  isotopes.  Based on the analysis by the
 antisymmetrized molecular dynamics, we show that the  $\alpha$  spectroscopic factor at nuclear
 exterior decreases in neutron-rich C isotopes, and the clustering suppression looks correlated
 with the growth of the neutron-skin thickness.
\end{abstract}

\section{Introduction} \label{sec:intro}
 It has been theoretically expected that the formation of $\alpha$ clusters in dilute nuclear
 matter is dependent not only on nuclear density, but also on the symmetry
 energy~\cite{Typel2010,Hagel2012}. Accordingly,  in finite nuclei, we also expect that the
 neutron-skin affects $\alpha$ clustering at nuclear exterior. A theoretical model
 study~\cite{Typel2014} has predicted  the  suppression of surface $\alpha$  clustering in
 neutron-rich nuclei due to the growth of  neutron skin.  
 
 Recently, the $\alpha$ knockout $(p,p\alpha)$  reaction has been established as a quantitative
 probe for the $\alpha$ clustering at nuclear
 surface~\cite{Yoshida2016,Wakasa2017,Yoshida2018,Yoshida2019}, and using this reaction,  the 
 $\alpha$ cluster  formation at the surface of Sn isotope chain has been measured for the first
 time~\cite{Tanaka2021}. It has been  shown that the knockout  cross section monotonically decreases in
 neutron-rich Sn isotopes indicating the negative correlation between  the neutron skin formation
 and $\alpha$ clustering. Hence, the growth of neutron skin looks preventing the $\alpha$ clustering
 at the surface of heavy nuclei. 

 Motivated by this fascinating discovery, we investigate the relationship between the neutron-skin
 thickness and $\alpha$ clustering in C isotopes. The advantages of studying light carbon isotopes
 is in reducing the theoretical uncertainty about nuclear clustering. The consistent description of
 $\alpha$ clusters and neutron skin within a same theoretical framework is, in  general, still
 challenging~\cite{Lovas1998,Qi2019}. However, there are ample theoretical and 
 experimental studies for C isotopes and its core nuclei (Be
 isotopes)~\cite{VonOertzen2006,Kanada-Enyo2012,Kimura2016}, and their 
 low-lying spectra have been established well. Furthermore, the decades of the studies have revealed
 the $\alpha$ clustering of  $^{12}{\rm C}$ in
 detail~\cite{Tohsaki2001,Funaki2003,Schuck2016,Zhou2016}. Thanks to these prior knowledge, 
 we are able to make a quantitative anatomy of the relationship between the neutron skin and
 $\alpha$ clustering for C isotopes.

This paper is organized as follows. In the next section, the theoretical framework of the
antisymmetrized molecular dynamics (AMD)~\cite{Kanada-Enyo2012,Kimura2016,Kanada-Enyo2003} used to
calculate the wave functions of Be and C isotopes, and the method to evaluate the $\alpha$
spectroscopic factors are briefly explained. In the section~\ref{sec:result}, we present the
numerical results and  discuss the relationship between the neutron-skin thickness and $\alpha$
clustering. It has been found that the $\alpha$ clustering of C isotopes likely show the negative  
correlation with the neutron-skin thickness similarly to Sn isotopes.  The final section summarizes
this work.

\section{Theoretical framework} \label{sec:framework}
\subsection{Hamiltonian and model wave function}
The microscopic $A$-body Hamiltonian is used in this study. It reads,
\begin{align}
 H=\sum_{i}^{A} t_{i} - t_{\rm cm} + \frac{1}{2}\sum_{ij}^{A} v_{\rm NN}(ij)
 +\frac{1}{2}\sum_{ij \in\rm {proton}}^{Z} v_{\rm C}(ij),
\end{align}
where the Gogny D1S parameter set~\cite{Berger1991} is employed as nuclear density functional 
$v_{\rm NN}(ij)$, and the Coulomb interaction $v_{\rm C}(ij)$ is approximated by a sum of seven
Gaussians.  The center-of-mass kinetic energy $t_{\rm cm}$ is subtracted from the total energy
without any approximation. 

The model wave function is a parity-projected Slater determinant,
\begin{align}
 \Phi^{\pi}=P^{\pi} \mathcal{A} \{\varphi_{1}\varphi_{2}\dots\varphi_{A}\},
\end{align}
where $P^\pi$ denotes parity projector, and $\varphi_{i}$ is the nucleon wave packet expressed by 
deformed Gaussian~\cite{Kimura2004a},
\begin{align}
\displaystyle \varphi_{i}(\bm r) &= 
\exp \set{-\sum_{\sigma=x,y,z} {\nu_{\sigma}} \left (r_{\sigma}-Z_{i\sigma} \right)^2 }
\chi_{i} \tau_{i},\\
\chi_{i}&=a_{i}\chi_\uparrow+b_{i}\chi_\downarrow, \quad
\tau_{i}= \{\rm proton\ or\ neutron\}.
\end{align}
The parameters of the model wave function are the Gaussian widths $(\nu_{x}, \nu_{y},
\nu_{z})$, centroids $\bm Z_i$, and the spin direction $a_i$ and $b_i$. They are determined by the 
frictional cooling method which minimizes the sum of the Hamiltonian and constraint potential,
\begin{align}
{E}(\beta)  &=\frac{\braket{\Phi^\pi|H|\Phi^\pi}}{\braket{\Phi^\pi|\Phi^\pi}} 
+v_{\beta}(\langle \beta \rangle - \beta)^2,
\end{align}
where the constraint potential strength $v_\beta$ is chosen sufficiently large value so that the
deformation of the model wave function $\braket{\beta}$~\cite{Kimura2012} is equal to the input
value $\beta$ by the minimization of $E(\beta)$. And we obtain the optimized wave function 
$\Phi^{\pi}(\beta)$ which has the minimum energy for each given value of $\beta$.

After the energy variation, the optimized wave functions are projected to the eigenstate of the
angular momentum, 
\begin{align}
 \Phi_{MK}^{J\pi}(\beta) = \frac{2J+1}{8\pi^2}\int \mathrm{d}\Omega
 D_{MK}^{J*}(\Omega)R(\Omega)\Phi^{\pi}(\beta),
\end{align}
where $D^{J}_{MK}(\Omega)$ and $R(\Omega)$ denote the Wigner's D-function and the rotation
operator. Then, the projected wave functions which have different value of the deformation
parameter $\beta$ are superposed as,
\begin{equation}
\Psi_{\alpha}^{J\pi}=\sum_{iK}  g_{iK\alpha} \Phi_{MK}^{J\pi} (\beta_{i}).\label{eq:gcmwf}
\end{equation}
Namely, Eq.~(\ref{eq:gcmwf}) is the wave function of the generator coordinate method
(GCM)~\cite{Hill1953} which employs the deformation parameter $\beta$ as the generator coordinate.  The
coefficients $g_{iK\alpha}$  and eigenenergy $E_{\alpha}$ are obtained by solving the Hill-Wheeler 
equation~\cite{Hill1953},   
\begin{align}
 &\sum_{j K^{\prime}}(H_{iKjK'}-E_\alpha N_{iKjK'})g_{j K'\alpha}=0\label{eq:gcm}\\
 &H_{iKjK'}=\braket{\Phi_{MK}^{J\pi}(\beta_{i})|H|\Phi_{MK'}^{J\pi}(\beta_{j})} ,\\
 &N_{iKjK'}=\braket{\Phi_{MK}^{J\pi}(\beta_{i})|\Phi_{MK'}^{J\pi}(\beta_{j})}.
\end{align}
From the GCM wave functions, we calculate the properties of C isotopes such as the
excitation spectra, proton and neutron distribution radii and electric transition probabilities.

\subsection{$\alpha$ reduced width amplitude}
To evaluate the degree of alpha clustering in C isotopes, we calculate $\alpha$ reduced width
amplitude (RWA) from the GCM wave functions. It is the probability amplitude to find $\alpha$
cluster at distance $a$ from the daughter nucleus, and is defined as 
\begin{align}
a y_\ell(a) = \sqrt{\tbinom{A}{4}}
 \braket{\delta(r-a)\Phi_\alpha[\Phi_{\rm Be(\ell^+)}Y_\ell(\hat r)]_0|\Phi_{\rm C}},
 \label{eq:rwa}
\end{align}
where $\Phi_\alpha$, $\Phi_{\rm Be(\ell^+)}$ and $\Phi_{\rm C}$ are the wave functions of $\alpha$ 
cluster,  the ground and excited states of Be isotopes with spin-parity $\ell^+$ and the ground
state of C isotopes, respectively. The wave functions of Be and C isotopes are calculated by GCM,
while that of $\alpha$ cluster is assumed to be the $(0s)^4$ configuration. Note that the angular
momentum $\ell$ of the orbital motion between $\alpha$ and Be isotope, and the spin of Be isotopes
which is also $\ell$ are coupled to the total angular momentum zero. In other words, we calculate
$\alpha$ RWA in the $\ell^+\times\ell=0^+\times  0$, $2^+\times 2$ and $4^+\times 4$ channels
$(\ell \leq 4)$, where the first number represents the spin of Be isotope and the second number
represents the orbital angular momentum between $\alpha$ and Be isotope.  In the practical numerical
calculation, Eq.~(\ref{eq:rwa}) has been evaluated by using the Laplace expansion
method~\cite{Chiba2017}. 

The degree of the $\alpha$ clustering may be evaluated by the so-called $\alpha$ spectroscopic factor
which is the squared integral of the $\alpha$ RWA,
\begin{align}
 S_\alpha(\ell^+\times \ell) = \int_0^\infty r^2 dr\ y_\ell^2(r).
\end{align}
It is noted that $S_\alpha$ is not normalized unity because of the antisymmetrized effects
between the $\alpha$ cluster and Be isotopes. As already discussed in the preceding
studies~\cite{Yoshida2018,Yoshida2019}, the $\alpha$ knockout reaction is sensitive only to the
$\alpha$ particle formed at the nuclear surface. Therefore, we expect that the integral of $\alpha$
RWA in nuclear exterior is a good measure for the knockout cross section. So, we introduce the
$\alpha$ spectroscopic factor integrated only in the exterior region,  
\begin{align}
 S^>_\alpha(\ell^+\times \ell) = \int_{\sqrt{\braket{r_m^2}}}^\infty r^2 dr\ y_\ell^2(r),
\end{align}
where $\sqrt{\braket{r_m^2}}$ denotes the root-mean-square radii of matter distribution of C
isotopes. 

\section{Results and Discussions}\label{sec:result}
\subsection{Structure of Be and C isotopes}
The structure of neutron-rich Be and C isotopes has already been studied in detail within the AMD
model and is summarized in the review papers~\cite{VonOertzen2006,Kimura2016}. Therefore, we do not
go into detail here, but only explain the important points necessary for the discussion of $\alpha$
clustering.  
\begin{table*}
\begin{center}
\caption{The deformation parameter $\beta$; the $E2$ transition probability; the proton, neutron and
 matter distribution radii; and the $\alpha$ spectroscopic factors for $\ell=0$ channel of the
 calculated ground states of Be and C isotopes. The $B(E2)$ values are given in the unit of $e^2\rm
 fm^4$, while the proton, neutron,  matter radii and neutron-skin thickness $\Delta r$ are given in
 the unit of fm.}\label{tab:tab1} 
\begin{tabular}{ccccccccc}
 \hline\noalign{\smallskip}
 &$\beta$ & $B(E2\uparrow)$ & $\sqrt{\braket{r_p^2}}$ & $\sqrt{\braket{r_n^2}}$ & $\sqrt{\braket{r_m^2}}$ & $\Delta r$ 
		 & $S_\alpha(0^+_1\times 0)$ & $S^>_\alpha(0^+_1\times 0)$ \\
\noalign{\smallskip}\hline\noalign{\smallskip}
 $^{10}{\rm Be}$ & 0.56 & 11.2 & 2.43 & 2.50 & 2.47 & 0.07 &      &     \\
 $^{12}{\rm Be}$ & 0.60 & 14.3 & 2.63 & 2.91 & 2.82 & 0.28 &      &     \\
 $^{14}{\rm Be}$ & 0.59 & 12.8 & 2.63 & 3.03 & 2.92 & 0.40 &      &     \\
 $^{12}{\rm C}$  & 0.50 & 11.2 & 2.52 & 2.52 & 2.52 & 0.00 & 0.30 & 0.24\\
 $^{14}{\rm C}$  & 0.34 & 1.4  & 2.54 & 2.59 & 2.57 & 0.05 & 0.10 & 0.08\\
 $^{16}{\rm C}$  & 0.39 & 5.6  & 2.60 & 2.83 & 2.74 & 0.23 & 0.05 & 0.04\\
 $^{18}{\rm C}$  & 0.45 & 4.7  & 2.65 & 2.98 & 2.87 & 0.33 & 0.04 & 0.04\\
\noalign{\smallskip}\hline
\end{tabular}
\end{center}
\end{table*}
\begin{figure*}[h]
 \begin{center}
  \includegraphics[width=0.7\textwidth]{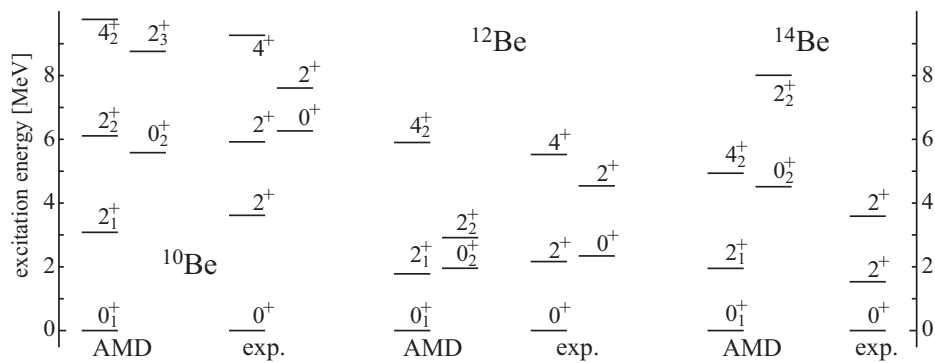}
  \caption{The calculated low-lying spectra of Be isotopes compared with the experimental
  data~\cite{Ajzenberg-Selove1991,Tilley2004,Kelley2017}. Only the positive-parity states are
  shown. } 
  \label{fig:levbe}       % Give a unique label
 \end{center}
\end{figure*}
\begin{figure}
 \begin{center}
  \includegraphics[width=0.45\textwidth]{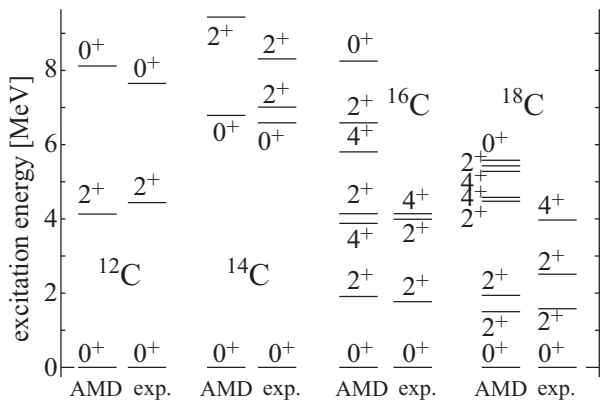}
  \caption{Same with Fig.~\ref{fig:levbe} but for C isotopes. Experimental data are taken from
  Refs.~\cite{Tilley2004,Kelley2017,Ajzenberg-Selove1986,Tilley1995}.} 
  \label{fig:levc}       % Give a unique label
 \end{center}
\end{figure}
\begin{figure}
 \begin{center}
  \includegraphics[width=0.45\textwidth]{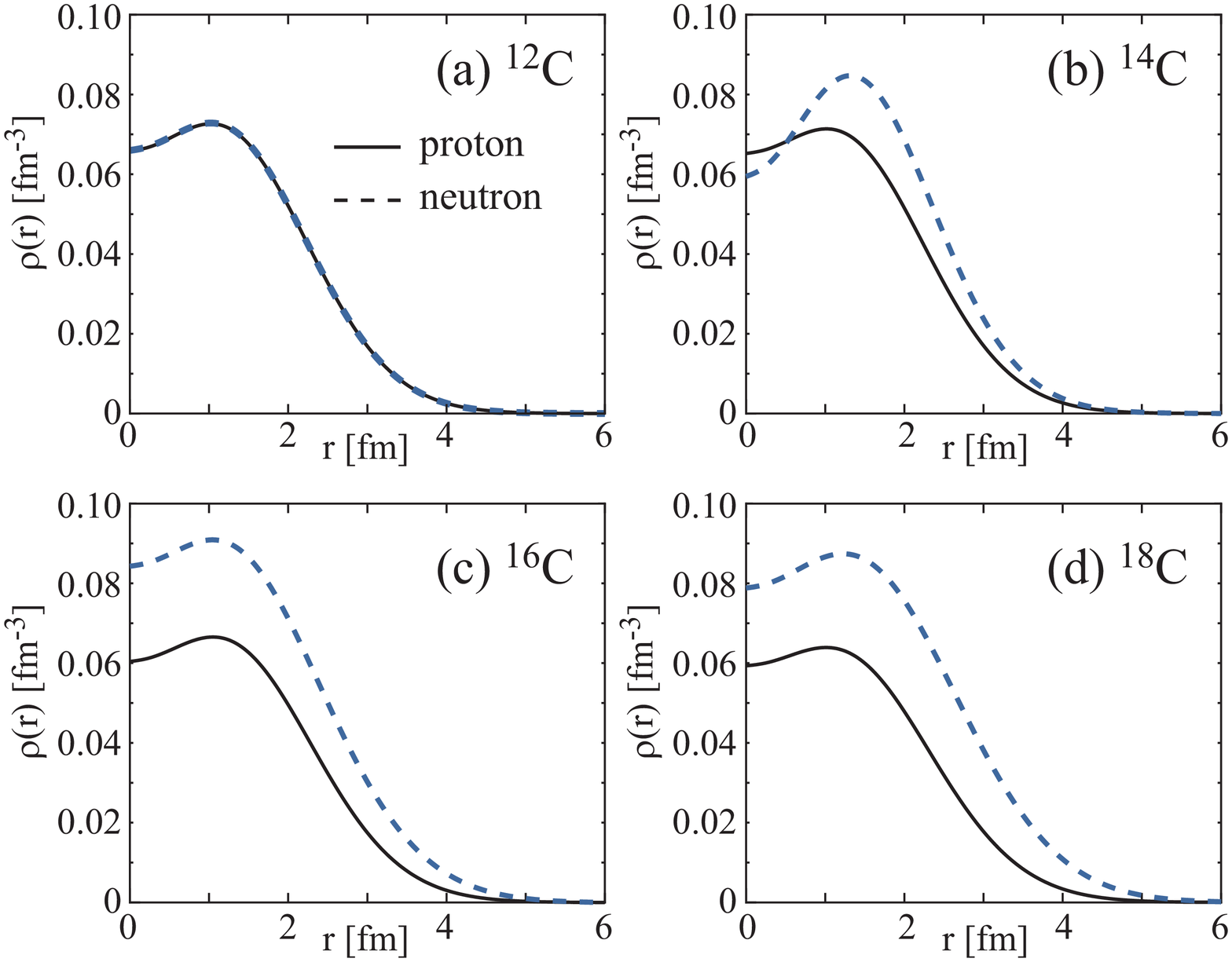}
  \caption{The calculated point proton and neutron density distributions of the ground states of C
  isotopes. The densities are normalized to the particle numbers. }
  \label{fig:density}       % Give a unique label
 \end{center}
\end{figure}
Firstly, we note that this is the first systematic calculation of both Be and C isotopes using
the Gogny D1S density functional, and it plausibly and consistently describes the low-lying spectra
as shown in Fig.~\ref{fig:levbe} and \ref{fig:levc}. This encouraging result ensures the reliability
of the analysis of the $\alpha$ cluster formation presented in the next section.   

As is well known, all calculated Be isotopes have dumbbell-shaped structure due to the pronounced
2$\alpha$ clustering~\cite{VonOertzen1996,Oertzen1997}. This brings about the strong deformation and $B(E2\uparrow)$ of the
ground states (Table.~\ref{tab:tab1}). If we investigate the property of individual isotope a little 
closer, we notice that $^{12}{\rm Be}$ and $^{14}{\rm Be}$ are more deformed than 
$^{10}{\rm Be}$. This is due to the difference of the valence neutron orbits in these
isotopes. Namely, two valence neutrons of $^{12}{\rm Be}$ and $^{14}{\rm Be}$ occupy the
$sd$-shell~\cite{VonOertzen2006}. On the other hand, all valence neutrons of $^{10}{\rm Be}$ occupy the $0p$
orbits which tend to reduce the inter-cluster distance. Note that our calculation reproduces the
well-known breaking of $N=8$ magic number~\cite{Talmi1960} in $^{12}{\rm Be}$ as two neutrons are
promoted into $sd$-shell from $p$-shell across $N=8$ shell gap. We also note that difference in the
valence  neutron configuration is clearly reflected to the neutron distribution radii
$\sqrt{\braket{r_n^2}}$. As a result, the neutron-skin thickness  
$\Delta r=\sqrt{\braket{r_n^2}} -\sqrt{\braket{r_p^2}}$ continuously increases toward the
neutron-drip-line nucleus $^{14}{\rm Be}$.   

We also comment on the non-yrast states of $^{10}{\rm Be}$ and $^{12}{\rm Be}$. $^{10}{\rm Be}$ has
the $2^+_2$ state approximately at 6 MeV which is reproduced our calculation. This state has the
same single-particle configuration with the ground state but originates in the triaxial deformation 
of its intrinsic structure~\cite{Itagaki2002}. On the other hand, the $0^+_2$ and $2^+_3$ states of
$^{10}{\rm Be}$ have the internal structure different from the ground state. Namely, these are the
neutron $2\hbar\omega$ excited states in which two neutrons are promoted into $sd$-shell across
$N=8$ shell gap. The opposite occurs in $^{12}{\rm Be}$ whose ground state is dominated by the
$2\hbar\omega$ configuration as already mentioned, but its non-yrast states (the $0^+_2$ and $2^+_2$
states) are dominated by the $0\hbar\omega$ state with the $N=8$ closed shell configuration. Thus,
the present calculation reasonably reproduces the known characteristics of the low-lying states of
Be isotopes.

\begin{figure*}
 \includegraphics[width=\textwidth]{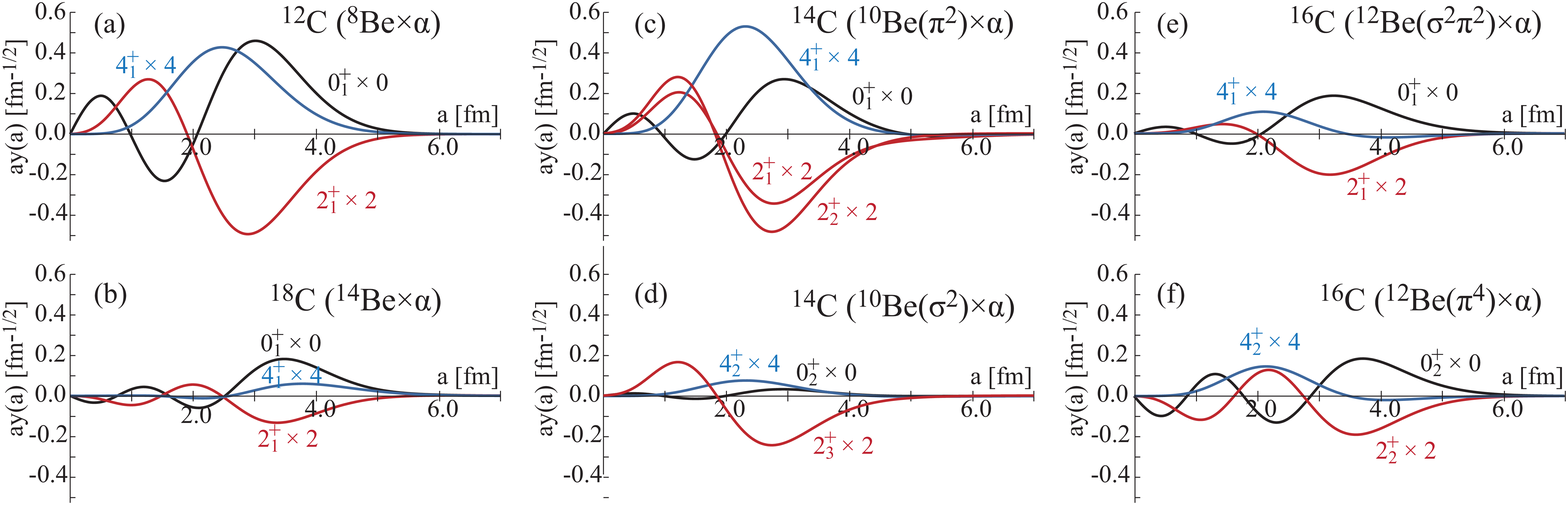}
 \caption{The calculated $\alpha$ RWA of carbon isotopes in the $\ell^+_n\times\ell$ channels, where
 $\ell^+_n$ denotes the spin-parity of $^{A}{\rm Be}$ while $\ell$ denotes orbital angular momentum
 between $\alpha$ and $^{A}{\rm Be}$. The panels (d) and (f) show the RWAs of $^{14}{\rm C}$ and
 $^{16}{\rm C}$ in the $\alpha+{}^{A}{\rm Be^*}$ channels where Be isotopes are excited to the
 non-yrast states. }  
 \label{fig:rwa}
\end{figure*}
\begin{figure}
 \begin{center}
  \includegraphics[width=0.40\textwidth]{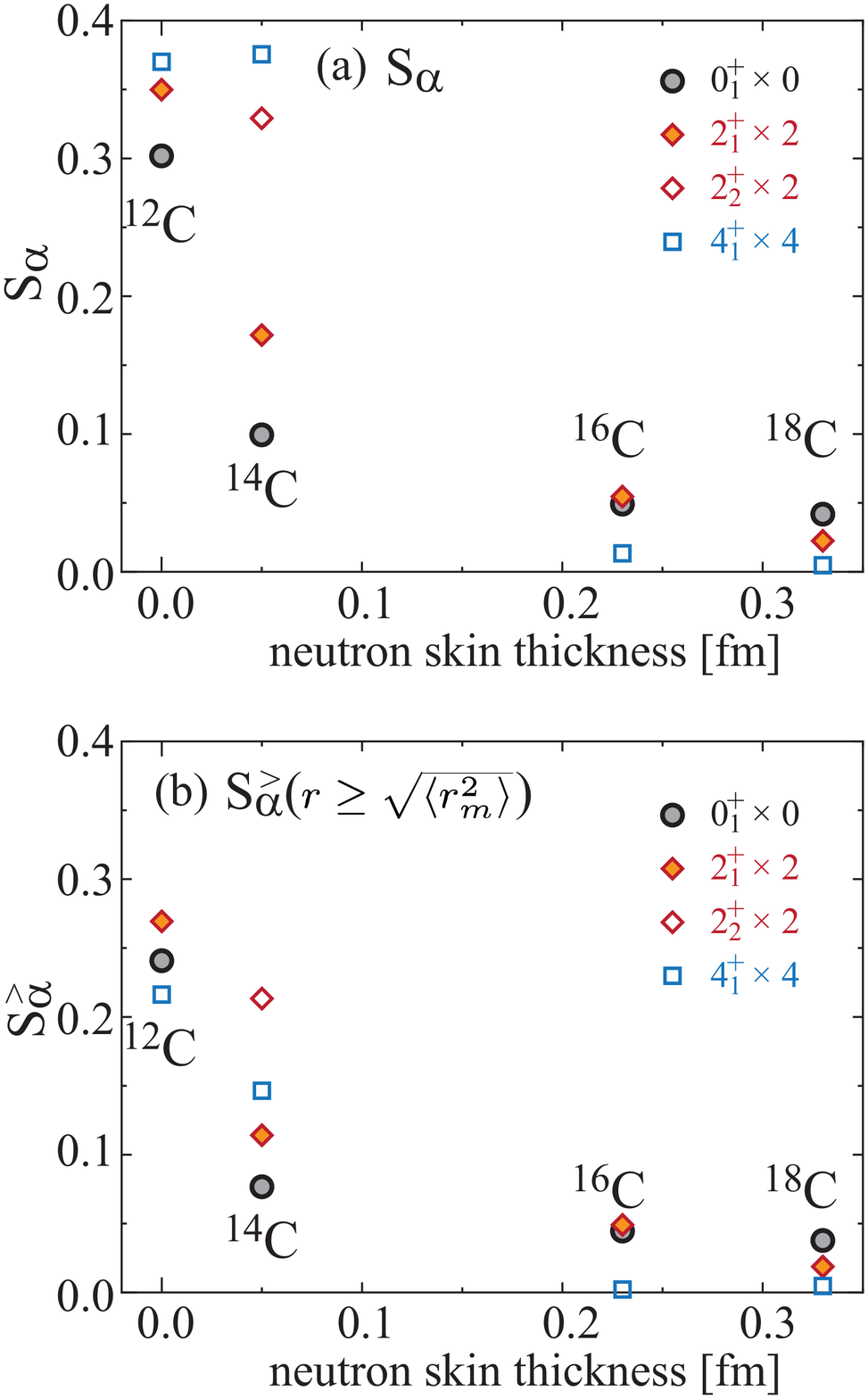}
  \caption{(a) The calculated $\alpha$ spectroscopic factors as function of the neutron-skin
  thickness. (b) Same with the panel (a), but the RWAs are integrated in the nuclear exterior
  ($r \geq \sqrt{\braket{r_m^2}}$). }
  \label{fig:RvsS}       % Give a unique label
 \end{center}
\end{figure}

Next, we discuss the spectra of C isotopes shown in Fig.~\ref{fig:levc}. Because of the
$Z=N=6$ sub-shell closure and $N=8$ shell closure, $^{12}{\rm C}$ and  $^{14}{\rm C}$ are rather
stiff and have large excitation energies of the $0^+$ and $2^+$ states. On the contrary, 
$^{16}{\rm C}$ and $^{18}{\rm C}$ are rather soft and have many low-lying states, as they
are weakly bound and have extra neutrons in an open shell ($sd$-shell). It is notable that the
present calculation reasonably described the excitation spectra of both stiff and soft
nuclei. Another interesting point to be noted is that all C isotopes except for $^{12}{\rm C}$ have
small $B(E2\uparrow)$ values despite their non-small quadrupole
deformation~\cite{Kimura2016,Kanada-Enyo2005,Jiang2020} as listed in Table.~\ref{tab:tab1}. This
result is consistent with the trend of the observed $E2$ 
transitions~\cite{Imai2004,Ong2008,Petri2012,Voss2012} and  indicates that the $2^+_1$ states of C
isotopes are dominated by neutron excitation not by proton excitation~\cite{Elekes2004}. 

The neutron single-particle configuration also affects the neutron-skin thickness
(Table.~\ref{tab:tab1} and Fig.~\ref{fig:density}). The neutron skin of $^{14}{\rm C}$ is
negligibly thin as two excess neutrons occupy the same major shell ($p$-shell) with protons. 
In $^{16}{\rm C}$ and $^{18}{\rm C}$, the excess neutrons start to occupy $sd$-shell, which pushes
the neutron density distributions outwards as seen in Fig.~\ref{fig:density}. Because of the
deformation, the valence neutron wave function is not a pure $d_{5/2}$ but an admixture of the
$d$- and $s$-wave components. This also helps to increase the neutron distribution
radii. Consequently,  $^{16}{\rm C}$ and $^{18}{\rm C}$ have much thicker neutron skin than
$^{14}{\rm C}$. The question we will examine below is how this neutron skin affects the $\alpha$
clustering at nuclear surface.   

\subsection{$\alpha$ formation probability and neutron-skin thickness}
Using the GCM wave functions explained above, we have evaluated the RWAs for the ground states of C
isotopes, which are shown in Fig.~\ref{fig:rwa}. 

As expected, the RWAs of the self-conjugate nucleus $^{12}{\rm C}$ (Fig.~\ref{fig:rwa}, panel
(a)) have the largest amplitude in all of the $\ell=0$, 2 and 4 channels. The amplitudes of $\ell=0$
and 2 are peaked at nuclear exterior ($r \geq 2.5$ fm), but oscillates and is suppressed in the
interior due to the Pauli exclusion. The $\ell=4$ amplitude peaks slightly inward because it has
centrifugal barrier and no Pauli forbidden state. $^{14}{\rm C}$ also has pronounced RWAs (panel
(b)) in the exterior region but they are slightly suppressed compared to $^{14}{\rm C}$. It is noted
that $^{10}{\rm Be}(2^+_2)$ state have almost the same intrinsic structure of $^{10}{\rm
Be}(0^+_1)$, and as a result, the RWA in the  $^{10}{\rm Be}(2^+_2)$ channel is also large. On the
other hand, the RWAs in other non-yrast state channels, {\it i.e.} $^{10}{\rm
Be}(0^+_2,2^+_3,4^+_2)$, are suppressed. This is due to the excited valence neutron configuration
of $^{10}{\rm Be}$.

The RWAs of $^{16}{\rm C}$ and $^{18}{\rm C}$ show quite different behavior. They are strongly
suppressed in both of the yrast (panels (b) and (e))  and the non-yrast (panel (f)) channels. 
Especially, the amplitudes are suppressed in the surface region where neutron skin exists. 
To quantitatively evaluate this suppression, Fig.~\ref{fig:RvsS} shows the $\alpha$ spectroscopic
factors integrated over whole space (panel (a)) and those integrated in nuclear exterior (panel (b))
as function of neutron-skin thickness. It looks that the reduction of the $\alpha$ clustering
correlates with the neutron-skin thickness, and this trend is consistent with the reduction of the
$(p,p\alpha)$ knockout reaction cross section observed in Sn isotopes~\cite{Tanaka2021}. Thus, the
$\alpha$ clustering suppression by the growth of the neutron-skin likely also exists in light mass
isotope chain.  

However, we also mention an alternative interpretation of Fig.~\ref{fig:rwa} and \ref{fig:RvsS}. As
we have already discussed, the neutron-skin thickness and the structure of the core nucleus (Be
isotopes) are strongly dependent on the neutron single-particle configuration. Therefore, the
neutron shell effect can be the real cause of the $\alpha$ clustering suppression observed in these
figures. We should also consider the relationship between the $\alpha$ clustering and $\alpha$
threshold energy~\cite{Ikeda1968}. To identify the real cause of the $\alpha$ clustering suppression, the
systematic study of other isotopes chains such as O, Ne and Mg isotopes is important, and the
research is now undergoing.

\section{Summary}
In this study, we have investigated the relationship between the neutron-skin thickness and $\alpha$
clustering of C isotopes to elucidate the possible clustering suppression by neutron skin. The AMD
framework has successfully described the low-lying spectra of both isotope chains
simultaneously. Using the obtained wave functions, we have evaluated the neutron-skin thickness and 
$\alpha$ clustering. It has been shown that $^{16}{\rm C}$  and $^{18}{\rm C}$  have thick neutron
skin, while $^{12}{\rm C}$ and $^{14}{\rm C}$ do not. The calculated $\alpha$ spectroscopic factors
show the negative correlation with the neutron-skin thickness. Namely, $\alpha$ clustering is
considerably suppressed in $^{16}{\rm C}$  and $^{18}{\rm C}$. Thus, the growth of the neutron skin
seems to suppress the $\alpha$ clustering of C isotopes similarly to those observed in Sn isotopes. 
However, we also point out that neutron shell effect may also play the crucial role and can be the
real cause of the $\alpha$ clustering suppression. This will be clarified by investigating the
trends in the neighboring isotopes chains.

\begin{acknowledgements}
One of the author (M.K.) acknowledges that this work was supported by the JSPS KAKENHI Grant No. 19K03859
and by the COREnet program at RCNP Osaka University. Part of the numerical calculations were
performed using Oakforest-PACS at the Center for Computational Sciences in the University of Tsukuba. 
\end{acknowledgements}

\bibliographystyle{spphys}       % APS-like style for physics
\bibliography{rwa.bib}   % name your BibTeX data base

% \begin{thebibliography}{}
% %
% % and use \bibitem to create references. Consult the Instructions
% % for authors for reference list style.
% % Format for Journal Reference
% Author, Article title, Journal, Volume, page numbers (year)
% Author, Book title, page numbers. Publisher, place (year)
% \end{thebibliography}

\end{document}